\documentclass[conference]{IEEEtran}
\IEEEoverridecommandlockouts
 
\usepackage{cite}
\usepackage{amsmath,amssymb,amsfonts}
\usepackage{algorithmic}
\usepackage{graphicx}
\usepackage{graphics}
\usepackage{textcomp}
\usepackage{xcolor} 
\usepackage{epsfig}
\usepackage{subcaption}
\usepackage[font=small,skip=0pt]{caption}
\usepackage{amsmath,amssymb,amsthm}
\usepackage{comment} 
\usepackage{tabularx}
\usepackage{graphicx}
\usepackage{enumitem}
\usepackage{eufrak}
\usepackage{lipsum}
\usepackage{stfloats}
\usepackage{bbm}
\usepackage{relsize}
\usepackage{multicol}
\usepackage[ruled]{algorithm2e}
\definecolor{darkred}{rgb}{0.1, 0.1, 0.60}

\newtheorem{propos}{Proposition}

\DeclareMathOperator{\erfc}{erfc}

\def\BibTeX{{\rm B\kern-.05em{\sc i\kern-.025em b}\kern-.08em
    T\kern-.1667em\lower.7ex\hbox{E}\kern-.125emX}}
\begin{document}

\title{
	 Latency-Optimal Cache-aided Multicast Streaming via Forward-Backward Reinforcement Learning
}

\author{
  \IEEEauthorblockN{Mohsen Amidzadeh}
	\IEEEauthorblockA{Department of Computer Science, Aalto University, Finland\\
	mohsen.amidzade@aalto.fi}	
}
\maketitle

\begin{abstract}
We consider a cellular network equipped with cache-enabled base-stations (BSs)
leveraging an orthogonal multipoint multicast (OMPMC) streaming scheme.
The network operates in a time-slotted fashion to serve content-requesting users by streaming cached files.
The users being unsatisfied by the multicat streaming face a delivery outage,
implying that they will remain interested in their preference at the next time-slot, which leads to a forward dynamics on the user preference.
To design a latency-optimal streaming policy, the dynamics of latency is properly modeled and included in the learning procedure.
We show that this dynamics surprisingly represents a backward dynamics.
The combination of problem's forward and backward dynamics then develops 
a forward-backward Markov decision process (FB-MDP) that fully captures the network evolution across time.
This FB-MDP necessitates usage of a forward-backward multi-objective reinforcement learning (FB-MORL) algorithm  
to optimize the expected latency as well as other performance metrics of interest including the overall outage probability and total resource consumption.
Simulation results show the merit of proposed FB-MORL algorithm in finding a promising dynamic cache policy.
\end{abstract}


\section{Introduction}\label{Sec_Intro}
Wireless caching is a promising approach to address the issues of data congestion and traffic escalation in cellular networks \cite{Bastug2014}.
In order to create an effective cache strategy, t
wo phases of cache placement and cache delivery\,/\,streaming need to be taken into account.

There are two main approaches for cache placement, probabilistic and deterministic. 
In contrast to the deterministic approach, the probabilistic placement can be scaled to large networks \cite{Zhou_2018,Cheng2019}. 
In this method, cache-equipped nodes randomly store files based on a network-wide common probability distribution.
This methodology is prevalent in cache-enabled policies 
within cellular and  wireless access networks \cite{Cheng2019,Wang2022,Zhang2024-6G}.

For the cache streaming, we utilize the  multipoint multicast (MPMC) scheme, 
which can provide more promising cache delivery
than conventional single-point unicast (SPUC) scheme for files with skewed popularity \cite{Amidzadeh2023,Guo2024multicast}.
MPMC involves multiple serving nodes broadcasting files cooperatively across the network, 
which makes it as a content-centric delivery scheme.
Notice that MPMC is in contrast to SPUC scheme that satisfies requesting User Equipments (UEs) 
individually by on-demand transmissions.
MPMC is prevalent in the literature and industry.
The Long Term Evolution (LTE) system incorporates Multipoint Multicast (MPMC) delivery 
to support the enhanced multimedia broadcast-multicast service (eMBMS) \cite{Militano2015}. 
An MPMC scheme also has been considered together with coded caching at the user end in \cite{Bayat2019}. 
Orthogonal MPMC streaming in a Single-Frequency-Network (SFN) configuration has been utilized 
in \cite{Amidzadeh2023} for edge caching cellular networks.
In \cite{Guo2024multicast}, an MPMC scheme is designed for an unmanned
aerial vehicle (UAV)-assisted cellular network.

In recent years, reinforcement learning (RL) has been widely used to design dynamic cache policies in diverse cellular networks \cite{Wei2022,Gu2022codedmulticast,Araf2023UAVcaching,Zhou2023EnergyOffloading,Qiong2024}.
In \cite{Wei2022}, an actor-critic RL algorithm is developed to obtain a proactive cache policy
optimizing the network metrics of caching cost and expected downloading delay.
In \cite{Gu2022codedmulticast}, the authors exploit a Policy Gradient (PG) RL algorithm to design a computation offloading policy for a cache-enabled network.
An actor-critic RL algorithm is leveraged in  \cite{Araf2023UAVcaching}
to design a cooperation cache policy for a  UAV-assisted two-tier cellular network.
The cooperation between aerial and ground BSs is then addressed to
optimize the global cache hit ratio.
A PG algorithm is used in \cite{Zhou2023EnergyOffloading} to design a service cache policy for the mobile edge computing (MEC)-enabled cellular networks.
In \cite{Qiong2024}, the authors propose a  multi-agent RL algorithm to 
design a cache placement in a cellular network consisting of a content server (CS) and caching BSs.

Latency is a paramount Quality-of-Experience (QoE) factor for designing an optimum streaming policy \cite{Sun2021,Amidzadeh2024,Zeng2024VideoStreaming,Liu2024VideoStreaming,Choi2025}.
In \cite{Amidzadeh2024}, the authors consider a multicast streaming for cache-enabled cellular networks 
with the transmission latency optimized using the harmonic broadcasting scheme. 
In \cite{Zeng2024VideoStreaming}, the latency is considered as a performance metric
in developing a video streaming policy for a cache-aided network with an edge node and cloud server.
In \cite{Liu2024VideoStreaming}, the latency is considered as a QoE metric for devising a video streaming  scheme
in a cache-enabled multi-tier cellular network. The authors apply an RL algorithm 
to jointly optimize the cache placement and user bit-rate to optimize the streaming scheme.
In \cite{Choi2025}, a latency-optimal video streaming is proposed using an RL algorithm for cache-aided MEC-enabled networks.
However, these research directions do not consider the effect of  transmission outage
in  analysing the  dynamics of the latency.
In contrast, we take into account this effect across different time-slots and show that it develops a backward dynamics 
that can be represented solely by a  backward MDP.
Consideration of this backward dynamics with the dynamics of user preference 
then provide a Forward-Backward Markov Decision Process (FB-MDP), a new class of MDPs \cite{amidzadehfbmoac}.
We then obtain an optimal dynamic caching by adopting a forward-backward RL algorithm \cite{amidzadehfbmoac}
on the basis of Advantage Actor-Critic (A2C) \cite{A2C}.
	
The contribution of this paper is listed as follows:
\begin{itemize}[leftmargin=*]
\item We design a latency-optimal cache-aided streaming by fully analyzing the dynamics of a content-centric OMPMC scheme.

\item We represent the problem based on a forward-backward Markov decision process (FB-MDP), 
which can exclusively model the time evolution of the network.

\item We leverage a forward-backward multi-objective RL approach built upon the A2C algorithm
to find an optimum multicast streaming taking into account resource usage, expected latency, and outage probability perspectives.

\end{itemize}
\section{Background} \label{Sec_background}

\subsection{Forward-Backward Markov Decision Process}
We here explain the notion of multi-objective forward-backward MDPs (FB-MDPs), 
expressed by a tuple 
$\left(\mathcal{S}, \mathcal{Y}, \mathcal{A}, P_f(\cdot), P_b(\cdot), \boldsymbol{r}^f(\cdot), \boldsymbol{r}^b(\cdot) \right)$,
where:
$\mathcal{S}$ and $\mathcal{Y}$ are the forward and backward state-spaces, respectively;
$\mathcal{A}$ is the action space;
$P_f\!: \mathcal{S}\times\mathcal{A}\times\mathcal{S}\to[0,1] $ is the forward transition probability describing the forward dynamics;
$P_b\!: \mathcal{Y}\times\mathcal{A}\times\mathcal{Y}\to[0,1]$  is the backward transition probability expressing the backward dynamics;
$\boldsymbol{r}^f:\mathcal{S}\times\mathcal{A}\to\mathbb{R}^{|S_f|}$ %
and $\boldsymbol{r}^b:\mathcal{Y}\times\mathcal{A}\to\mathbb{R}^{|S_b|}$ are the forward and backward reward functions
(respectively), where $S_f$ and $S_b$ are the sets of indices of the forward and backward rewards (respectively). %
The forward transition probability describes the incremental evolution of forward state
based on the current state $\textbf{s}_t \in \mathcal{S}$ and action $\textbf{a}_t \in \mathcal{A}$.
Moreover, in a time-reversed way, the previous backward state of the system follows $\textbf{y}_{t-1}\sim  P_b(\cdot|\textbf{y}_t,\textbf{a}_t)$ 
from $\textbf{y}_t \in \mathcal{Y}$ by performing action $\textbf{a}_t \in \mathcal{A}$. %

We here need to stress that a FB-MDP cannot be expressed as standard MDP \cite{amidzadehfbmoac}.
The aim of a FB-MDP problem is thus to optimize the following discounted multi-objective cumulative reward from the Pareto-optimality perspective:
\begingroup\makeatletter\def\f@size{8.5}\check@mathfonts
\begin{align}\label{Eq:FBMDP}
	\!\!\!\max_{\{\textbf{a}_t \in \mathcal{A}\}_{t\in \{1,T\}}} \:\mathbb{E} \left\{ \sum_{t=1}^T \gamma^{t-1} \:\Big[\boldsymbol{r}^f(\textbf{s}_t, \textbf{a}_t),~\boldsymbol{r}^b(\textbf{y}_{T-t+1}, \textbf{a}_{T-t+1}) \Big] \right\},
\end{align}
\endgroup
where $T \in \mathbb{N}$ is the finite horizon,
$\gamma \in [0,1]$ the discount factor,
and the expectation refers to the different realizations of the forward-backward trajectories.

\subsection{RL Algorithm for Solving Multi-Objective FB-MDPs}\label{sec:solution-characteristics}
Solving problem in \eqref{Eq:FBMDP} requires a multi-objective RL algorithm
so as to find a policy distribution for the action, i.e.,  $\textbf{a}_t \sim \pi(\cdot | \textbf{s}_t)$ 
that can simultaneously learn both the forward and backward dynamics.
This forward-backward RL algorithm leverages a step-wise chronological mechanism including three main phases: (i) \textit{forward pass}, in which 
the forward dynamics is simulated by generating actions using the policy $\textbf{a}_t \sim \pi(\cdot|\textbf{s}_t)$; 
(ii) \textit{backward pass}, in which
the backward dynamics is simulated in a time-reversed way 
by leveraging the actions generated in the previous step;
and (iii) \textit{bidirectional learning}, which employs a multi-objective optimization mechanism \textit{with a suitable chronological order} to optimize the policy $\pi(\cdot|\textbf{s}_t)$
based on the experiences obtained from both the forward and backward dynamics. 
In this paper, we utilize the FB-MOAC algorithm \cite{amidzadehfbmoac} built upon the aforementioned mechanism
for our cache strategy design.

\section{System Model} \label{Sec_Model}
We consider a cellular network with cache-enabled base-stations (BS).
We use a Poisson Point Process (PPP) $\Phi_{\rm bs}$ with intensity $\lambda_{\rm bs}$ to model the deployment of BSs.
The network operates in a time-slotted fashion indexed by 
$t\in \{1,\ldots,T\}$,
where $T$ stands for finite time horizon.
At each time-slot, there exists a set of UEs that prefer files from a content library.
Without loss of generality, we assume the library contains $N$ different files with the same length equal to L.
These segments are behaved as distinct entities.
Note that for the case of files with different length, 
this assumption can be eliminated by partitioning the files into smaller segments of equal lengths \cite{Rezaei2022}.
Contents have different popularity $\{ p_n^{\rm pop}(t) \}_{n=1}^N$, 
where $p_n^{\rm pop}(t)$ is the network-wide popularity for file $n$ 
indicating the probability that content $n$ is requested by a randomly selected user at time-slot $t$. 
The goal is to satisfy as many users as possible during the network operation.

\begin{figure}[t]
	\centering
	\includegraphics[width=6 cm]{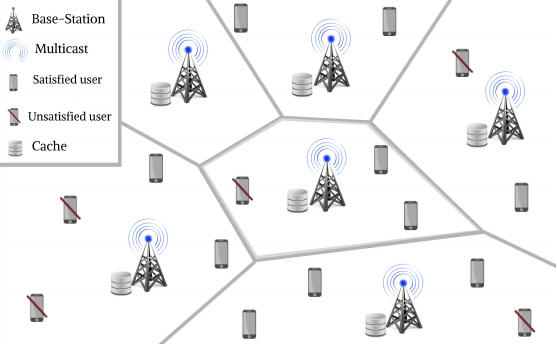}
	\captionof{figure}{The multicast content delivery scheme. 
			The BSs collaborate	with each other to stream the cached files towards users.
		It is probable that some users becomes un-satsified due to multicast outage. \vspace{-10 pt}}
	\label{Fig_Hybrid}
\end{figure}

The network serves the UEs by the OMPMC streaming scheme employing BSs, as depicted in Figure \ref{Fig_Hybrid}.
For this, the BSs apply OMPMC to cooperatively broadcast the cached files across the whole network.
OMPMC exploits file-specific disjoint radio resources to eliminate the interference during streaming different files.
As a consequence, the broadcast scheme of BSs constitutes a content-centric network \cite{Ours_2022_twc}.
Note that some UEs being served by OMPMC component get dissatisfied due to the transmission outage.

\subsection{Cache Placements}
In the multicast layer, BSs have limited cache capacity which allows them to store $C$ files at most. 
They utilize a probabilistic cache placement policy to store files, as described in reference \cite{Blaszczyszyn2015}. 
For this, a network-wide file-specific distribution $\{p_n^{\rm cach}(t)\}_{n=1}^N$ is used,
where $p_n^{\rm cach}(t) \in [0,1]$ denotes the probability file $n$ is cached in a randomly selected BS at time-slot $t$.
In order to adhere to the cache capacity, the sum of all cache probabilities must be equal $C$, i.e., $\sum_{n=1}^N p_n^{\rm cach}(t) = C$.


\subsection{OMPMC Streaming}
In each time-slot of network operation,
there exists a spatial distribution of UEs preferring files.
The network responds to these UEs by applying the OMPMC scheme.
The OMPMC component thus streams cached files across the network by cooperation of all BSs.
It exploits file-specific disjoint resources $\{w_n(t)\}_{n=1}^N$ to broadcast different files, 
where  $w_n(t)$ is the bandwidth allocated for file $n$ at time-slot $t$. 
To reduce the latency of multicast streaming, the harmonic broadcasting (HB) \cite{Juhn1997} is incorporated in OMPMC. 
HB is characterized by a time-varying harmonic number $N_{\rm hb}(t)$ and works as follows.
The expected latency experienced by a typical UE at time-slot $t$ is reduced by a factor of $1/M(t)$,
if the transmission bandwidth is increased by a factor of $N_{\rm hb}(t) = \sum_{i=1}^{M(t)} 1/i$. 
It means that the UE does not need to wait for the entire duration of a broadcasted file to be able to download it from the beginning.
Rather, it can wait for a fraction of $1/M(t)$ of the file duration, on average.
This considerably reduces the latency by efficiently increasing the bandwidth.
Therefore, the duration of time-slot $t$ is $d(t) = \frac{L}{M(t)}$ seconds, where $L$ is the length of the broadcasted file in seconds.
Note that $d(t)$ can also be translated as the latency experienced by a UE to start receiving the broadcasted file,
and we call it time-slot resolution.
For instance, when the harmonic number is set to $N_{\rm hb}(t)=7$, the bandwidth is increased by a factor of $7$. 
In this case, we obtain $M(t)=620$, since $\sum_{i=1}^{620} \tfrac{1}{i} \approx 7$ \cite{Juhn1997}. 
Consequently, the expected latency for a file of one hour in duration 
is effectively reduced to $\tfrac{1}{2}\tfrac{3600}{620} \approx 3$ seconds.

For all BSs, we assume the same average transmission power denoted by $p_{\rm tx}$. 
We apply a power allocation scheme
with average power $p_{\rm tx} \frac{w_n}{W}$ being used to stream file $n$.
As such, the transmitting Signal-to-Noise-Ratio (SNR) of all files in OMPMC are the same, 
$$ 
\gamma_{\rm tx} = \frac{p_{\rm tx} w_n(t)/W}{w_n(t) N_0} = \frac{p_{\rm tx} }{W N_0},
$$
where $N_0$ is the noise spectral density.
With file-specific resource allocation of OMPMC, the Signal-to-Noise-Ratio of UE $k$ receiving file $n$ is expressed as \cite{Ours_2020}
\begin{equation} \label{EQ_SNR_MPMC}
	\gamma_{k,n} = \gamma_{\rm tx} \sum_{j\in\Phi_{\rm bs,n}}{ |h_{j,k}|^2 \|\boldsymbol{x}_k-\boldsymbol{r}_j \|^{-e}},
\end{equation}
where a standard distance-dependent is used to model the path-loss with  $e$ the path-loss exponent.
Moreover, $\Phi_{\rm bs,n}$ is the set of BSs caching file $n$, 
$h_{j,k}$ is the channel coefficient between BS $j$ and UE $k$ with a Rayleigh distribution, i.e., $|h_{j,k}|^2\sim \exp(1)$ 
and $\boldsymbol{x}_k$ and $\boldsymbol{r}_j$ are the locations of UE $k$ and BS $j$, respectively.
We now evaluate the outage probability of OMPMC component that is translated to the
probability that a typical UE being served by OMPMC cannot decode the broadcasted file.
The capacity of Additive-White-Gaussian-Noise channel gives the maximum achievable transmission rate.
If this rate experienced by a UE is less than the minimum required rate $R$, the UE is in outage. 
Therefore, the outage probability $\mathcal{O}_{n,k}$ for UE $k$ receiving file $n$ from OMPMC is:
$$
\mathcal{O}_{n,k}(t) = \mathbb{P}\{ w_n(t)\: \log_2(1+\gamma_{k,n}) \leq R \}.
$$
We now define a spectral efficiency $\alpha_n(t) = R/w_n(t)$.
As such, the total resource consumption of OMPMC component is 
$W(t)= N_{hb}(t) \sum_{n=1}^N w_n(t) = N_{hb}(t)  \sum_{n=1}^N \frac{R}{\alpha_n(t)} $,
where $N_{hb}(t)$ is added due to the HB scheme.
The outage probability can be computed for a typical UE located at the origin, based on the Slivnyak-Mecke theorem \cite{BK1}. 
By setting $\mathcal{O}_{n,0} = \mathcal{O}_{n}$,
the outage probability for broadcasted file $n$ with path-loss exponent $e=4$ is thus \cite{Ours_2020}:
\begingroup\makeatletter\def\f@size{9}\check@mathfonts
$$
\mathcal{O}_n(t) = \erfc\bigg( \frac{\pi^2 \lambda_{\rm bs} p_n^{\rm cach}(t)}{4} \sqrt{\frac{\gamma_R}{\eta_n(t)}} \bigg).
$$
\endgroup
where \scalebox{0.9}{$\gamma_R = \dfrac{p_{\rm tx}}{N_0} \dfrac{1}{R\:N_{\rm hb}(t)}$ and $\eta_n(t)= \big( 2^{\alpha_n(t)}-1 \big)\sum_{n=1}^N 1/\alpha_n(t)$}.

\subsection{File Popularity and Intensity of UE Request}
The dynamics of UE intensity requesting a specific file depends on the file popularity
and  the success of content streaming scheme.
Specifically, certain users fail to receive the requested content in the current timeslot due to the outage probability; 
Their request is thus deferred to the subsequent one.
Hence, each time-slot sees a distribution of users accounting for the repeated requests 
and a distribution describing the new preferences toward contents.
This leads to a time-varying model for the request probability of content $n$, $p_n^{\rm req}(t)$, described as follows.
\begin{propos}
	Consider the OMPMC streaming scheme serving users that request $N$ contents with network-wide file popularities $\{p_n^{\rm pop}(t)\}_{n=1}^N$
	and experiencing the outage probability $\{\mathcal{O}_n(t)\}_{n=1}^N$.
	Then, the dynamics of the request probability of files complies with the following forward dynamics.
	\begingroup\makeatletter\def\f@size{9}\check@mathfonts
	\begin{align}\label{Eq:F-MDP}
	\!\!\!\! {\color{darkred} p_n^{\rm req}(t)} =& \underbrace{ {\color{darkred} p_n^{\rm req}(t-1)} \mathcal{O}_n( t-1)}_{\text{repeated request}} \notag\\
	&+ \underbrace{p_n^{\rm pop}(t)\!\!\sum_{m=1}^N \big( 1- \mathcal{O}_m( t-1)\big) {\color{darkred} p_m^{\rm req}(t-1)} }_{\text{new request based on the popularity}}.
	\end{align}
	\endgroup
\begin{proof}
	Please refer to Appendix of the pre-print version.
\end{proof}
\end{propos}

Equation \eqref{Eq:F-MDP} 
illustrates that at each time-slot, there are two distinct source for requesting UEs: 
one requesting files based on the popularity $p_n^{\rm pop}(t)$, and another one repeating their previous request due to the presence of an outage.

\subsection{Expected Latency for Successful Delivery}
Considering that a file request might be repeated several time-slots until successful reception,
we intend to express the expected latency required to successfully receive file $n$ at time-slot $t$.
Let $L_n(t)$ denote it, we then get:
\begin{propos}
	Consider the OMPMC streaming scheme operating in time-slots with duration $d(t)$ and under the outage probability $\{\mathcal{O}_n(t)\}_{n=1}^N$.
	Then, the dynamics of the expected latency required to successfully receive file $n$ is described based on the following backward dynamics
	\begingroup\makeatletter\def\f@size{9}\check@mathfonts
	\begin{align}\label{Eq:B-MDP}
	\!\!\!	{\color{darkred} L_n(t) } = \mathcal{O}_n(t)\Big( d(t) + {\color{darkred} L_n(t+1) } \Big) + \big(1 - \mathcal{O}_n(t)\big)\frac12 d(t), 
	\end{align}	
	\endgroup
	where $~~L_n(T)=0$.
	\begin{proof}
		Please refer to Appendix of the pre-print version.
	\end{proof}
\end{propos}

Notice that we have $L_n(T)=0$ since system operations finish at time $t=T$ 
and the users do not need to wait any longer.
Eq. \eqref{Eq:B-MDP} represents a \textbf{backward dynamics}, with the backward state $L_n(t)$.
Note that this model fully captures the effect of outage probability in analyzing the evolution of latency and differs from the conventional models \cite{Sun2021,Wei2022,Yun2022,Zeng2024VideoStreaming,Liu2024VideoStreaming,Choi2025} that do not consider 
the impact of the outage in latency when accounting for successive slots; 
for the delivery without outage, the expected latency simply becomes $L_n(t)=\frac{d(t)}2$, as its realizations follow a uniform distribution with values between $0$ and $d(t)$.
Eq. \eqref{Eq:B-MDP} may suggest that it is possible to convert it to a standard forward dynamics.
For this purpose, one can consider a variable transformation $K_n(T-t) := L_n(t)$ 
as well as a time transformation $t':=T-t$ in order to obtain 
the following forward dynamics on $K_n(t')$:
\begingroup\makeatletter\def\f@size{9}\check@mathfonts
\begin{align*}
	K_n(t') =& \big( d(T-t') + K_n(t'-1) \big)\mathcal{O}_n(T-t')\\ 
	&+ \frac{d(T-t')}2 \big(1-\mathcal{O}_n(T-t')\big),~~\text{for}~t'\geq 1,
\end{align*}	
\endgroup
with $K_n'(0)=0$.
However, this shows a non-causal MDP, as the state $K_n(t')$ depends on the far future of outage $\mathcal{O}_n(T-t')$ 
that cannot be revealed by moving forward in time.
Eq. \eqref{Eq:B-MDP} also shows that
for a full-error streaming scheme (i.e., with the outage equal to one) $L_n(t) = d(t) + L_n(t+1)$ holds, 
which means that the expected latency maximally accumulates as one goes backwards in time.
This is expected, as no successful receptions take place.
Moreover, it is worth stressing that minimizing the expected latency in \eqref{Eq:B-MDP} enables 
to \emph{optimally} keep track of the \emph{precise} time-slot at which requests are finally fulfilled. %
Alternatively, one could track the service time of requests to prioritize those that have waited longer, or track for the failed\,/\,succeeded content transmissions.
However, these policies do not completely map to the tracking of overall latency,
and oversimplify the problem.
Consequently, they fail to account for the complex interactions within the system, leading to a sub-optimal solution.
The evaluation in Section \ref{Sec_Result} empirically confirms this claim. %

\section{Optimal Dynamic Caching}\label{Sec_Formul}
\subsection{Problem Modeling}
We here intend to design a dynamic  cache-aided streaming by considering multiple network performance metric, including the overal latency,
and modeling the problem based on a FB-MDP.
For this, we consider a finite time-interval of network operations $t\in[1,T]$.
Then, as a measure of network QoS, we take into account the overall probability of unsatisfied UEs.
This metric is expressed by
$
r_{\rm QoS}(t)=  \sum_{n=1}^N p_n^{\rm req}(t) {\mathcal{O}}_n(t).
$
Furthermore, we consider the total resource consumption of the OMPMC scheme:
$
r_{\rm BW}(t) = W_{\rm eff}(t) = N_{hb}(t) \sum_{n=1}^N w_n(t).
$	
We finally consider a cost related to the overall latency of hybrid delivery:
$
r_{\rm Lat}(t) = \sum_{n=1}^N p_n^{\rm req}(t) L_n(t),
$	
with $L_n(t)$ being obtained by \eqref{Eq:B-MDP}.
We now consider OMPMC parameters including 
the cache placement probabilities $\{p_n^{\rm cach}(t) \in [0,1]\}_{n=1}^N$,
spectral efficiencies $\{\alpha_n(t) \in [0,\infty)\}_{n=1}^N$ with $\alpha_n=\frac{R}{w_n}$
and harmonic number $N_{hb}(t) \in \mathbb{N}$,
as the action: $\textbf{a}(t) = [\{p_n^{\rm cach}\}_n, \{\alpha_n\}_n, N_{hb}](t)$.
Accordingly, the dynamics~\eqref{Eq:F-MDP} is cast as a forward MDP 
with forward state $\textbf{s}(t) = [\{p_n^{\rm req}(t)\}_n]$ and action $\textbf{a}(t)$ which affects the outage.
Moreover, the latency dynamics~\eqref{Eq:B-MDP} is cast as a backward MDP 
with backward state $\textbf{y}(t) = [\{L_n(t)\}_n]$. 
These two MDPs thus construct a FB-MDP with the forward rewards $[r_{\rm QoS}, r_{BW}](t)$,
and backward reward $r_{\rm Lat}(t)$.

\subsection{Streaming Policy Learning}
\begin{figure}[t]
	\centering
	\includegraphics[width=6 cm]{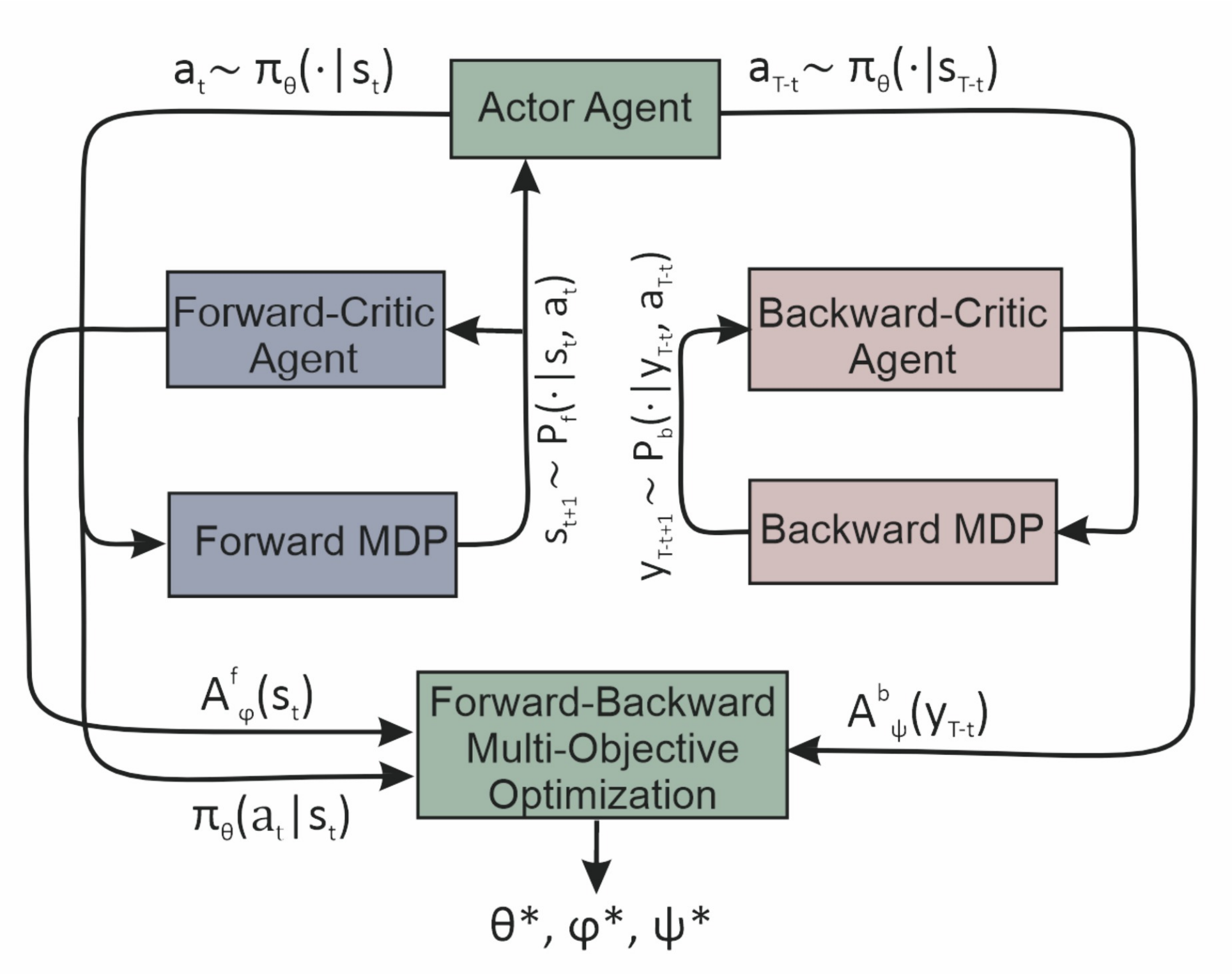}
	\captionof{figure}{Diagram of the forward-backward multi-objective RL. \vspace{-10 pt}}
	\label{Fig_FwBwRL}
\end{figure}

We aim to design a dynamic cache-aided streaming by optimizing cumulative summations of performance metrics $(r_{\rm QOS}, r_{\rm BW}, r_{\rm Lat})$ through time-slots $t\in\{1,T\}$.
This optimization can be formulated as an constrained maximization problem based on the following multi-objective cumulative reward function:
\begingroup\makeatletter\def\f@size{9}\check@mathfonts
\begin{align*}
	\mbox{O}_1: \!\!\!
	&\underset{
		\begin{small} 
			\begin{array}{c} p_n^{\rm cach}, \alpha_n, N_{\rm hb}  
			\end{array} 
		\end{small}}
	\max\!\!\!\! -\sum_{t=1}^T \gamma^{t-1} \Big[r_{\rm QoS}(t), r_{\rm BW}(t),  r_{\rm Lat}(T-t) \Big]\\
	&~~~\text{s.t.}
	\begin{cases}
		\text{FB-MDP in } \eqref{Eq:F-MDP} \text{ and } \eqref{Eq:B-MDP},&\\
		\sum_{n=1}^N p_n^{\rm cach}(t)= C, & 0\leq p^{\rm cach}_n(t)\leq 1, \\
		\alpha_n(t) \geq 0, \\
		N_{\rm hb}(t) \in \mathbb{N}.
	\end{cases}
\end{align*}
\endgroup
We then exploit the FB-MOAC algorithm \cite{amidzadehfbmoac} which is developed
for learning FB-MDP problems.
This RL algorithm is built based on a actor-critic architecture,
represented by NNs.
The single policy actor is parameterized by a $\theta$-parametric NN
to provide the policy distribution $\pi_\theta(\cdot | \textbf{s}_t)$.
However, apart from the actor network , 
it additionally includes forward and backward critic networks,
that adjust the actor network based on the simulations of forward and backward dynamics, respectively.
They are parameterized by two NNs with parameters $\phi$ and $\psi$,
and are criticizing the actor using the forward and backward advantage functions, $A^f_\phi(\textbf{s}_t)$ and $A^b_\psi(\textbf{y}_{T_t})$.
It then Pareto-optimize the reward functions based on a multi-objective optimization mechanism such that the expected-value of rewards can monotonically improve. 
Fig. \ref{Fig_FwBwRL} show the diagram of the FB-MOAC algorithm.

\section{Simulation Results and Discussion}\label{Sec_Result}

\subsection{Experiment Setup and Algorithm Parameters}
We follow the settings of \cite{Amidzadeh2023} for the considered environment. 
Specifically, the number of contents is set to $N=200$, 
the capacity of BSs to $C=10$,
the spatial intensity of BSs to $\lambda_{\rm bs}=100$ points/km$^2$,
and the 
transmission rate to $1$ Mbps,
and content length to $L=600$ seconds. 
The total number of time slots is $T = 256$.

We model the network-wide file popularity $p_n^{\rm pop}(t)$ 
by a diffusion model \cite{Srinivasan2024}, 
which provides a set of time-varying Zipf distributions with skewness $0.6$.
We apply an Urban NLOS scenario from 3GPP \cite{3GPP2017} with carrier frequency 2 GHz, HN transmission power 23
dBm, and path-loss exponent $e=4$. 
The antenna gains at the BSs are 8 dBi, the noise-figure of UE is 9 dB, the noise spectrum density is -174 dBm. 
The reference distance is 1 km, so the PPP intensities are in the units of points/km$^2$.
	
Three separate sets of NNs is considered for the actor, forward-critic and backward-critic networks in  \textit{FB-MOAC} algorithm.
The rectified linear unit (ReLU) activation function is used for the neurons connection.
The number of neurons in the hidden layer for the actor and critics is $100$,
the actor and forward\,/\,backward critic learning rates are $3\times10^{-4}$, and the smoothing factor to $\gamma_{\rm mov}=0.95$.

\subsection{Performance Evaluation}
\begin{figure*}[t!]
	\centering
	\subcaptionbox{\label{Fig:TrainMulticast1_APP}}{%
		\includegraphics[width=0.3\textwidth]{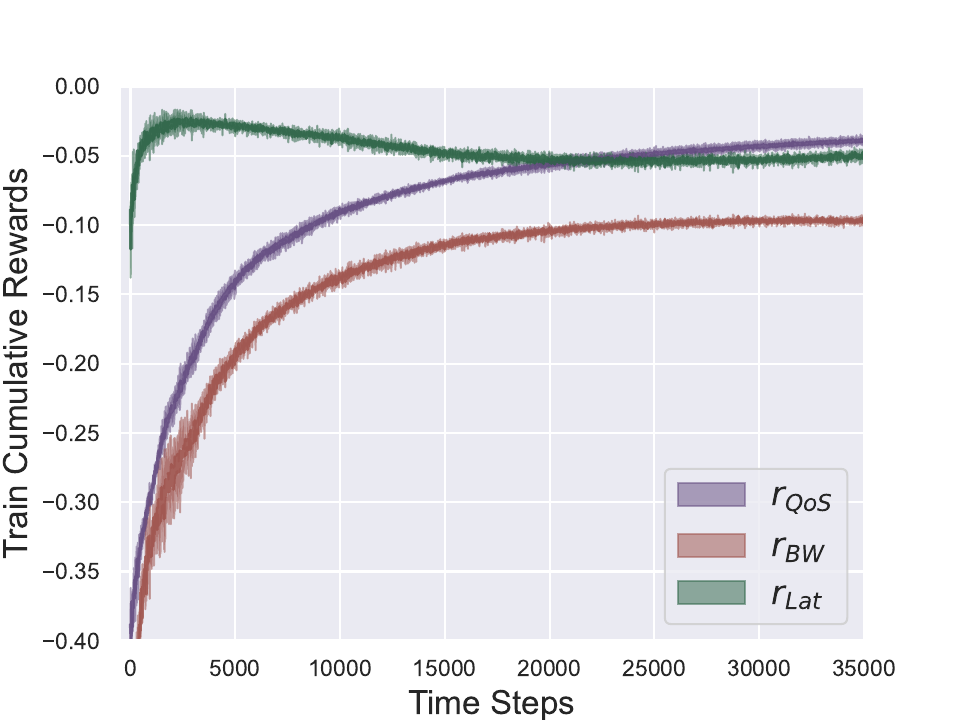}
	}%
	\subcaptionbox{\label{Fig:TrainMulticast2_APP}}{%
		\includegraphics[width=0.3\textwidth]{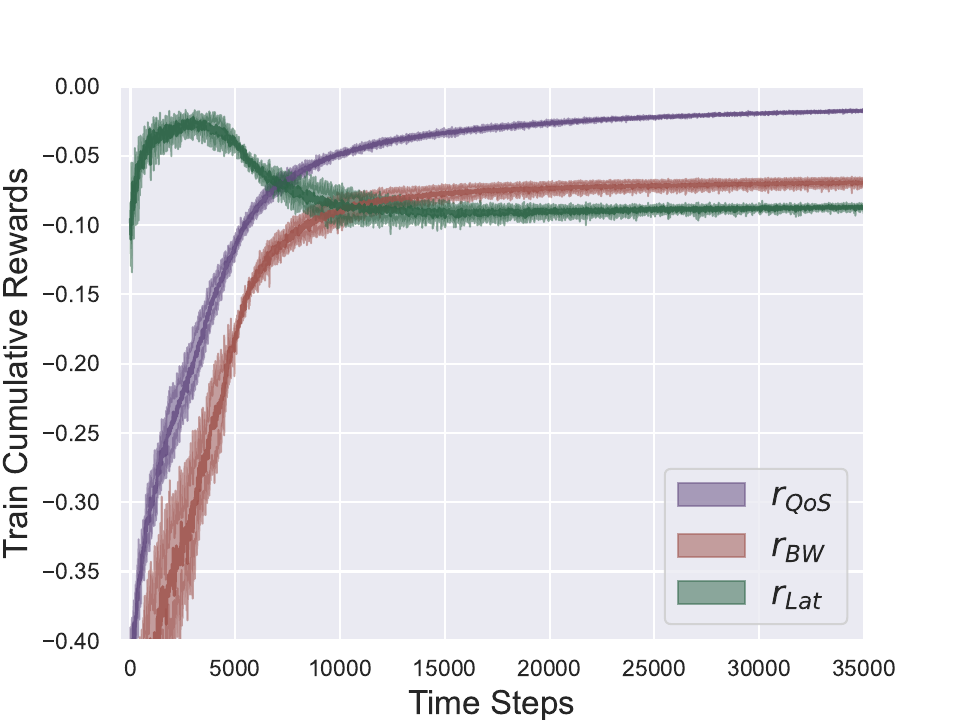}
	}%
	\subcaptionbox{\label{Fig:TrainMulticast3_APP}}{%
		\includegraphics[width=0.3\textwidth]{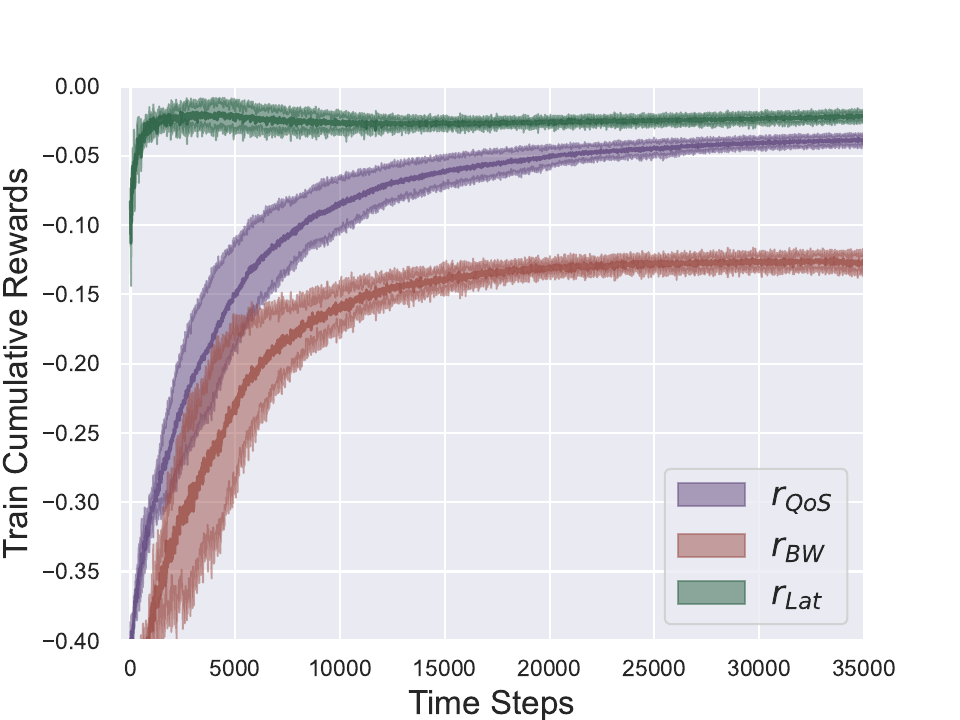}
	}%
	\hspace{-.5em}%
	\caption{Pareto-optimal solutions for  the cache-aided streaming policy with different preference settings on the considered reward functions,
	$[\alpha_{\rm QoS},\alpha_{\rm BW},\alpha_{\rm Lat}]$
	 (a) $[0.3, 0.3,1.0]$, 
	 (b) $[1.0,1.0,0.3]$, 
	 and (c) $[0.3,1.0,0.3]$. \vspace{-10 pt}
	}\label{fig:edge-caching:learning}
\end{figure*}
\begin{figure}[t]
	\centering
	\includegraphics[width=0.37\textwidth]{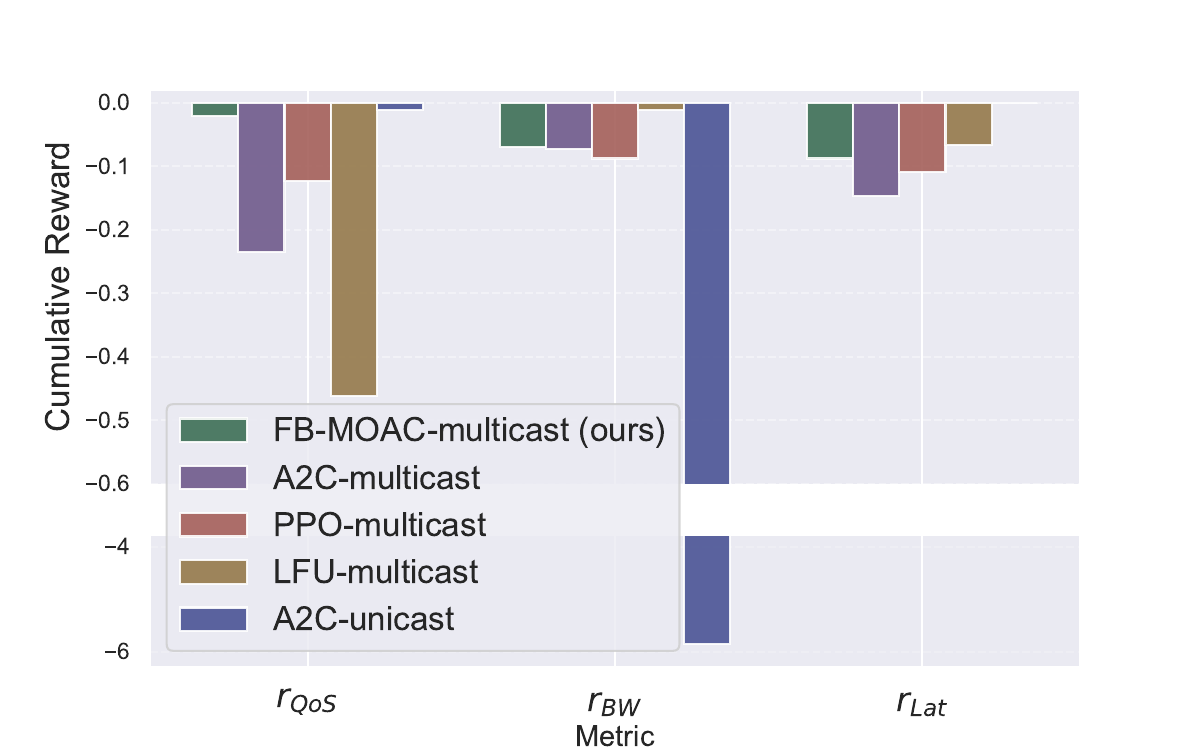}
	\caption{Comparison of multicast streaming obtained by FB-MOAC 
		against (i)  a multicast streaming obtained by rule-based LFU,
		(ii) an unicast streaming obtained by MOAC,
		(iii-iv) two multicast streaming policies obtained by forward-only PPO and MOAC algorithms.
		\label{Fig:CacheComparison_APP}
		\vspace{-10 pt}}
\end{figure}

Fig. \ref{fig:edge-caching:learning} shows the learned Pareto-optimal
solutions of proposed streaming  scheme obtained by FB-MOAC algorithm
with different preference settings  
applied on considered  reward function 
(Note that Pareto-optimal solutions are not unique, and we use  different  preference settings to potentially obtain  most of Pareto solutions \cite{amidzadehfbmoac}).
Since the respective solution of each figure does not dominate that of the others,
it has been able to obtain most of the Pareto-optimal solutions.
For clarity, the performance metrics are \textcolor{black}{normalized} with respect to $r_{\rm QoS}$, making them be presented together in a single plot.
All of the considered rewards 
are converged into a stable solution,
thereby the FB-MOAC algorithm is effectively learned.

To benchmark the streaming solution obtained by the FB-MOAC algorithm (termed as \emph{FB-MOAC-multicast}),
we consider four baselines:
(i) a multicast streaming policy based on the widely used rule-based Least Frequently Used (LFU) method~\cite{LFU} (denoted as \emph{LFU-multicast});
(ii) a learning-based unicast streaming approach with all contents available, 
which serves as a conventional benchmark in cellular networks \cite{Chakareski2024,andrews2016primer};
and (iii–iv) two learning-based multicast streaming approaches that optimally exclude the backward-MDP component \cite{Amidzadeh2024}.
For the unicast streaming, we use a multi-objective A2C (MOAC) algorithm
to optimize QoS and the bandwidth consumption,
and term the resulting solution as \emph{MOAC-unicast}.
For the latter learning-based methods, 
we  use this fact that optimizing $r_{\rm QoS}$ and $d(t)$ 
reduce $r_{\rm Lat}$ based on \eqref{Eq:B-MDP},
thereby we consider $r_{\rm QoS}$ and $r_{\rm BW}$ as forward rewards, 
and replace the backward reward with optimizing  $d(t)$.
We then use baseline RL algorithms  PPO~\cite{PPO2017} and MOAC to learn solution policies.
We term the resulting solutions of these strategies as \emph{PPO-multicast} and \emph{MOAC-multicast}, respectively.
Figure~\ref{Fig:CacheComparison_APP} compares the \emph{FB-MOAC-multicast} against baselines, in terms of normalized rewards.
For the \emph{FB-MOAC-multicast} and \emph{MOAC-unicast}, 
we select a solution among different Pareto solutions by prioritizing $r_{\rm QoS}$,
and for the \emph{PPO-multicast} and \emph{MOAC-multicast}, we learn forward rewards and optimize 
$d(t)$ to obtain a solution with  $r_{\rm Lat}$ comparable to that of \emph{FB-MOAC-multicast}.
For the unicast streaming baseline, note that the latency is zero, as the request
are immediately responded based on a on-demand delivery.

The results show that \emph{FB-MOAC-multicast} outperforms \emph{PPO-multicast} and \emph{MOAC-multicast} in all rewards,
which implies that FB-MOAC can provide a creditable multicast streaming scheme 
notably better than forward-only strategies.
Specifically, more than 15\% of the contents will be lost due to the values of 
QoS for \emph{PPO-multicast} and \emph{MOAC-multicast},
whereas less than \textbf{5\%} of them fails in \emph{FB-MOAC-multicast}.
Moreover, the cache policy of \emph{FB-MOAC-multicast} Pareto-dominates those of \emph{PPO-multicast} and \emph{MOAC-multicast}.
Although, the unicast streaming strategy has slightly better QoS than \emph{FB-MOAC-multicast},
it has been obtain at the cost of extreme bandwidth consumption;
bandwidth consumption of \emph{FB-MOAC-multicast} and \emph{MOAC-unicast} are, $3.3$ GHz and \textbf{$40$} GHz, respectively.
On the other hand, the LFU streaming is better than \emph{FB-MOAC-multicast} from the bandwidth-consumption perspective, 
however, it is remarkably unreliable because around \textbf{50\%} of the streams fail on this schemes.

The results show that multicast streaming can be considered as 
a promising candidate for dynamic settings in cellular networks compared to the conventional unicast, 
as it can remarkably reduce the bandwidth consumption with the same level of QOS of unicast 
and acceptable streaming latency.
Moreover, they show  the efficiency of our RL-based mechanism
in obtaining a dynamic solution for the  cache-aided network,
compared to the other rule-based and learning-based alternatives.

\section{Conclusion}\label{Sec_Concl}
In this paper, we considered a cache-enabled content streaming based on orthogonal multipoint multicast (OMPMC) scheme.
We then regarded time evolution of the network and aimed to find a latency-optimum streaming solution
with minimum resource usage and quality-of-service.
We found that our streaming  problem can be formulated exclusively based on a forward-backward Markov decision process (FB-MDP).
In order to obtain a solution for the formulated FB-MDP and tackle simultaneously multiple performance metrics, 
we leveraged a forward-backward multi-objective reinforcement learning (FB-MORL) algorithm.
The results showed the merit of FB-MORL in finding a promising solution.
We then benchmarked the performance of our dynamic cache delivery compared to other rule-based and learning-based alternatives. 
Simulation results show that our scheme significantly outperforms the conventional other approaches by a considerable margin. 
These findings indicate that the proposed dynamic  policy holds great promise as a cache-aided streaming scheme
and be leveraged with modern mobile networks that have an  eMBMS component.

As an interesting future work, we consider the combination of unicast and multicast cache-aided deliveries 
to develop a dynamics streaming scheme.
\bibliographystyle{IEEEtran}
\bibliography{IEEEabrv,IEEE}


\onecolumn
\appendix
\noindent \textbf{Proof of Proposition 1:}

Consider a randomly selected UE and define the events
\[
E_n(t) := \{\text{UE prefers file }n\text{ at time }t\},\qquad n \in \{1,\dots,N\},
\]
and
\[
S(t) := \{\text{UE is satisfied at time }t \text{ by OMPMC streaming} \},\qquad F(t):=S(t)^c.
\]
We also write $q_n(t)=\mathbb{P}\big(E_n(t)\big)$ and
\[
p_n(t)=\mathbb{P}\big(E_n(t)\mid S(t-1)\big),
\]
so that \(p_n(t)\) is the file popularity under an error-free (no-outage) network model.  We assume the sequence \(p_n(t)\) is known a priori (e.g. modeled by a time-varying Zipf profile \cite{Zipf}).

To obtain the file popularity \(q_n(t)\) under the proposed error-prone hybrid scheme, apply the law of total probability and condition on the file requested in the previous slot:
\begin{align*}
	q_n(t)
	&= \mathbb{P}\big(E_n(t)\big) \\
	&= \mathbb{P}\big(E_n(t)\mid S(t-1)\big)\,\mathbb{P}\big(S(t-1)\big)
	+ \mathbb{P}\big(E_n(t)\mid F(t-1)\big)\,\mathbb{P}\big(F(t-1)\big)\\[4pt]
	&= p_n(t)\sum_{m=1}^N \mathbb{P}\big(S(t-1)\mid E_m(t-1)\big)\,\mathbb{P}\big(E_m(t-1)\big)\\
	&\qquad\;+\sum_{m=1}^N \mathbb{P}\big(E_n(t)\mid F(t-1),E_m(t-1)\big)\,
	\mathbb{P}\big(E_m(t-1)\big)\,\mathbb{P}\big(F(t-1)\mid E_m(t-1)\big).
\end{align*}

Now introduce the file-dependent outage probability. Let
\[
\mathcal{O}_m(t-1) \;:=\; \mathbb{P}\big(F(t-1)\mid E_m(t-1)\big)
\]
be the total outage probability when file \(m\) was requested at \(t-1\). Then
\[
\mathbb{P}\big(S(t-1)\mid E_m(t-1)\big)=1-\mathcal{O}_m(t-1),
\qquad \mathbb{P}\big(E_m(t-1)\big)=q_m(t-1).
\]

We further assume that an unsatisfied UE that requested file \(m\) at \(t-1\) will request the same file again at time \(t\) with probability one, i.e.
\[
\mathbb{P}\big(E_n(t)\mid F(t-1),E_m(t-1)\big)=\mathbf{1}\{n=m\}.
\]
Using these identities the previous expression simplifies to
\begin{align}\label{EQ_FwDMP0_revised}
	q_n(t)
	&= p_n(t)\sum_{m=1}^N \big(1-\mathcal{O}_m(t-1)\big)\,q_m(t-1)
	+ q_n(t-1)\,\mathcal{O}_n(t-1),
\end{align}
for \(n=1,\dots,N\). Note that summing \eqref{EQ_FwDMP0_revised} over \(n\) yields \(\sum_n q_n(t)=\sum_n q_n(t-1)=1\), so the equation preserves probability mass.

\vspace{20 pt}

\noindent \textbf{Proof of Proposition 2:}

Define the following events for a UE requesting file \(n\) at time \(t\):
\[
S_n(t) := \{\text{UE is satisfied by OMPMC for file }n\text{ at time }t\},
\qquad
F_n(t) := S_n(t)^{c} = \{\text{delivery fails (outage) for file }n\text{ at }t\}.
\]
By definition the outage probability satisfies
\[
\mathbb{P}\big(F_n(t)\big) = \mathcal{O}_n(t), \qquad
\mathbb{P}\big(S_n(t)\big) = 1-\mathcal{O}_n(t).
\]

Let \(L_n(t)\) denote the expected latency to successfully receive file \(n\) starting at slot \(t\):
\[
L_n(t)=\mathbb{E}\big[\text{latency to successfully receive file }n\big].
\]
Conditioning on whether the UE is satisfied in the current slot or not gives
\begin{align*}
	L_n(t)
	&= \mathbb{E}\big[\text{latency}\mid S_n(t)\big]\,\mathbb{P}\big(S_n(t)\big)
	+ \mathbb{E}\big[\text{latency}\mid F_n(t)\big]\,\mathbb{P}\big(F_n(t)\big).
\end{align*}

Now use the (standard) slot-level latency assumptions:
\begin{itemize}
	\item If the delivery succeeds in the current slot (\(S_n(t)\)), the expected latency incurred in that slot is \(\tfrac{1}{2}d(t)\) (the success occurs on average halfway through the slot).
	\item If the delivery fails (\(F_n(t)\)), the UE waits the remainder of the slot (duration \(d(t)\)) and remains interested in the same file in the next slot; therefore the conditional expected latency is \(d(t) + L_n(t+1)\).
\end{itemize}
Hence
\begin{align*}
	L_n(t)
	&= \tfrac{1}{2}d(t)\,\big(1-\mathcal{O}_n(t)\big) + \big(d(t) + L_n(t+1)\big)\,\mathcal{O}_n(t).
\end{align*}
Rearranging yields the recursion
\begin{align}\label{EQ_BwMDP_events}
	L_n(t) \;=\; \mathcal{O}_n(t)\big(d(t) + L_n(t+1)\big) + \big(1-\mathcal{O}_n(t)\big)\tfrac{1}{2}d(t),
	\qquad L_n(T)=0,
\end{align}
where the terminal condition \(L_n(T)=0\) indicates no further latency is incurred after the final slot.

\end{document}